# Electrical Control of Magnetization in Charge-ordered Multiferroic $LuFe_2O_4$

Chang-Hui Li, Fen Wang, Yi Liu, Xiang-Qun Zhang, Zhao-Hua Cheng, and Young Sun*

Beijing National Laboratory for Condensed Matter Physics, Institute of Physics,

Chinese Academy of Sciences, Beijing 100190, P. R. China

Abstract

$LuFe_2O_4$ exhibits multiferroicity due to charge order on a frustrated triangular lattice. We find that the magnetization of $LuFe_2O_4$ in the multiferroic state can be electrically controlled by applying voltage pulses. Depending on with or without magnetic fields, the magnetization can be electrically switched up or down. We have excluded thermal heating effect and attributed this electrical control of magnetization to an intrinsic magnetoelectric coupling in response to the electrical breakdown of charge ordering. Our findings open up a new route toward electrical control of magnetization.

*Corresponding author. Electronic address: youngsun@aphy.iphy.ac.cn

The demand for higher data density in magnetic information storage has triggered an intense search for approaches to control magnetization by means other than magnetic fields [1-3]. For example, magnetization reversal by spin-polarized electrical currents through torque is a promising approach and under intense investigations [4,5]. The magnetoelectric effects - the influence of polarization by a magnetic field and of magnetization by an electric field - provide another possible way for magnetization manipulation [6,7]. However, compared with the wide observations of magnetic control of polarization in a variety of multiferroic materials [8-10], electrical control of magnetization is more difficult to achieve and rarely reported [11,12], especially in single-phase materials. In this Letter, we demonstrate giant changes of magnetization induced by applying voltage pulses in multiferroic $LuFe_2O_4$.

The mixed valence compound $LuFe_2O_4$ has recently received a lot of attention due to the charge order (CO) induced ferroelectricity and multiferroicity [13,14]. The compound has a double layer structure with a triangular iron lattice. The coulombic interactions between $Fe^{2+}$ and $Fe^{3+}$ ions compete with the frustrated nature of the triangular lattice, which leads to a peculiar ordered arrangement of charges. A three-dimensional (3D) CO occurs below 320 K [15,16], which results in a net electrical polarization. This kind of ferroelectricity associated with CO is termed as "electronic ferroelectricity" [13], in contrast to conventional ferroelectricity involving displacement of cation and anion pairs. Meanwhile, the strong magnetic interactions between Fe moments develop as a ferrimagnetic ordering below a Neel temperature $T_N \sim$ 240 K [17]. Therefore, $LuFe_2O_4$ is multiferroic below 240 K. Moreover, recent observations of the magnetocapacitance [18] and the sharp change of electrical polarization at $T_N$ [13], as well as the X-ray scattering study [19] indicate a magnetoelectric (ME) coupling in $LuFe_2O_4$.

Previous studies have shown that the CO in many systems can be broken down by external stimulus like electric fields [20,21]. Our recent work suggests that the CO in $LuFe_2O_4$ is very sensitive to external electric fields [22]. Considering the multiferroicity and the possible coupling between CO and magnetism in $LuFe_2O_4$, the breakdown of CO could induce a remarkable change in magnetism. To testify this interesting issue, we have performed a careful study by direct magnetization measurements with applied voltage pulses on $LuFe_2O_4$ single crystals. The results demonstrate spectacularly that the magnetization has dramatic changes in response to the breakdown of CO.

Single crystals of $LuFe_2O_4$ were grown by optical floating-zone melting method in a flowing argon atmosphere. X–ray diffraction at room temperature showed that the samples are single phase and have a structure consistent with literature. The $LuFe_2O_4$ crystal for the study has a size of 0.4 mm ×1.58 mm×0.7 mm (ab plane×c). The magnetization and I-V curves were simultaneously measured with a Quantum Design Superconducting Quantum Interference Device (SQUID) magnetometer using a probe constructed in-house. A Keithley 2400 source meter was used to supply dc voltage and measure the current.

Figure 1 shows the magnetization of a $LuFe_2O_4$ single crystal as a function of applied voltage at 200 K, along with the I-V curves measured in the same scanning round. The measurements were performed by scanning voltage pulse (~ 20 ms) using the configuration shown in the inset of Fig. 1(b). The magnetic field is applied along c axis (the easy magnetization direction) while the voltage pulse is applied in the ab plane where the CO is easier to be broken down than the c-axis direction. The current is recorded when the voltage pulse is on while the magnetization is measured after each pulse is off. The current limit is set to 0.1 A to avoid a burst of big current. The I-V curves measured in zero and 0.1 T are nearly the same and exhibit a big jump of current at a critical voltage (~ 22 V), consistent with our previous report on polycrystalline samples [22]. As observed in other CO systems such as $Pr_{1-x}Ca_xMnO_3$ manganites [20] and magnetite [21], the sudden jump of current indicates the breakdown of CO by applied electric field.

The magnetization as a function of applied voltage is shown in Fig. 1(c). The curve marked with H=0 T was measured as following: a 5 T magnetic field was applied along c axis to magnetize the sample, then the field is removed and the remanent magnetization was measured in zero magnetic field. The curve marked with H=0.1 T was measured in a constant 0.1 T field. With the increase of applied voltage, the magnetization changes smoothly when the voltage is below 17 V. This initial change could be the normal magnetic relaxation with time. When the applied voltage is above 17 V, the magnetization changes much faster and exhibit giant jumps at the critical voltage. In zero magnetic field, the remanent magnetization jumps down. In contrast, the magnetization jumps up in a constant 0.1 T magnetic field. We note that the magnetization change is independent of the polarity of the voltage. Both positive and negative voltages have the same effect. The above results imply that the breakdown of CO in $LuFe_2O_4$ leads to a remarkable change in magnetization.

To further confirm the electrically driven magnetization change, we performed a series of magnetization measurements with single voltage pulses at different temperatures and magnetic fields. Fig. 2(a) shows the results at 200 K. In zero magnetic field, the remanent magnetization is immediately switched to zero after applying a single 22 V voltage pulse. In a constant 0.1 T magnetic field, the magnetization is immediately switched up after applying a 22 V voltage pulse and remains in the high state. The successive 22 V voltage pulses have negligible influence on the magnetization. Fig. 2(b) shows the magnetization switch at 100 K. The I-V curve at 100 K suggests that the critical voltage has increased to ∼ 100 V. In zero magnetic field, the remanent magnetization is immediately switched down after applying a 100 V voltage pulse. In a constant 5 T magnetic field, the magnetization is immediately switched up after applying a 100 V voltage pulse. A little different from the case at 200 K, the successive second and third voltage pulses cause minor changes on the magnetization.

We must emphasize that this immediate and giant magnetization switch by voltage pulse can not be simply accounted for by thermal heating effects. As the applied voltage (22 V) generates a current of $10^{-2}$ A flowing through an area of $0.7 \times 1.58$ mm$^2$, the current density is only in the order of $10^0$ A/cm$^2$. Within the short pulse period (~ 0.02 s), the total thermal energy generated by the current is about 4.4 mJ. With the known specific heat of LuFe$_2$O$_4$ [23,24] and the mass of the sample (3.5 mg), the thermal energy would at most cause a sample temperature change of 3 K at 200 K. In order to further clarify the thermal effect, we directly monitored the temperature of the sample using a Pt temperature sensor glued with the sample. At 200 K, the sample temperature increases slightly (< 3 K) and then restores quickly within a couple of seconds when a 22 V voltage pulse is applied. Moreover, a smaller voltage pulse (18 V) causes a negligible temperature change but the magnetization already shows remarkable changes as seen in Fig. 1(c). One may argue that local thermal heating due to conduction filaments in the sample could happen and cause a change of magnetization. However, the magnetization of filaments should take only a small proportion of the total magnetization and certainly can not account for the huge change of magnetization in this case. Therefore, we conclude that the giant magnetization change induced by voltage pulses is an intrinsic property of the material.

Figure 3 shows the temperature dependence of c-axis magnetization in 0.1 and 5 T with the zero-field-cooling (ZFC) and field-cooling (FC) processes. There is a big divergence between the ZFC and FC magnetization below a freezing temperature. With increasing magnetic field, the freezing temperature and the starting point of the divergence shift to lower temperature. When we plot the initial and final magnetization states shown in Fig. 2 on these M-T curves, we find a clear relation between them. The electrical switch-up of magnetization in a magnetic field is actually a switch from the ZFC to the FC magnetization, as indicated by arrows in Fig. 3. In addition, we also measured the magnetization in a constant 5 T field at 200 K and found that the magnetization remains unchanged after applying a 22 V voltage pulse. Moreover, the measurement at 300 K shows no magnetization change by voltage pulse although the electrical breakdown of CO is easier at higher temperatures. Based on these results, we conclude that the electrical switch-down of remanent magnetization can always happen in the ferrimagnetic state whereas the electrical switch-up of magnetization in a magnetic field can only happen in the region where the ZFC and FC magnetization has a divergence. This relation indicates that the role of applied voltage pulse is to promote the relaxation process toward the equilibrium state. The physics process could be like the following: when a high enough voltage is turned on, the CO is broken down and the localized carriers are released. The motion of electrons drives a magnetic excitation through a ME coupling effect to overcome the energy barriers and fall immediately into the lowest energy valley. When the voltage is off, the CO is restored whereas the magnetic state is restrained in the valley.

In practical applications, a reversible switch between low and high magnetization states is usually required. In Fig. 4, we demonstrate the reversible switch of magnetic states at 200 K by employing the electrical control of magnetization. In state 1, the magnetization is set by applying a 0.1 T field; then the magnetization is switched up to a high level (state 2) by applying a voltage pulse; once the magnetic field is cut off the magnetization drops to a remanent level (state 3); finally, the magnetization is switched down to a very low level (state 4) by applying another voltage pulse. As shown in Fig. 5, these four states can be repeatedly reproduced for many times.

The electrical switch of magnetization can also be used for manipulating the magnetization hysteresis loop. Fig. 5 shows the magnetization hysteresis loops measured at 100 K after a ZFC process. The normal hysteresis loop shows a large coercive field of 2.5 T. When a 100 V voltage pulse is applied at -1 T in the descending branch, the magnetization switches from positive to negative (not to zero). Similarly, when a 100 V voltage pulse is applied at +1 T in the ascending branch, the magnetization switches from negative to positive. This result demonstrates that the magnetization reversal process can be electrically assisted.

Our study suggests that the electrical breakdown of CO in $LuFe_2O_4$ can induce immediate and giant changes in magnetization, which opens up a new route toward electrical control of magnetization. Similar effects could be expected in other materials with CO induced multiferrroicity. The physics of this electrical switch of magnetization could be related to an intrinsic ME effect through the coupling between the CO and magnetism. With the benefit of small critical electric fields required for CO breakdown and the ferrimagnetic rather than antiferromagnetic nature in $LuFe_2O_4$, this new strategy has appealing advantages over previously known approaches using spin-polarized/unpolarized current or electric fields, where a high current density [3,5] of $10^5$-$10^8$ $A/cm^2$ or a high electric field [11,25] of $10^5$-$10^6$ V/cm is usually required. In comparison, the magnetization switch in $LuFe_2O_4$ only requires a low current density of $10^0$ $A/cm^2$ or a small electric field of $10^2$-$10^3$ V/cm. Since the Neel temperature $T_N$ (~ 240 K) of $LuFe_2O_4$ is still below room temperature, future efforts would be devoted to the search for materials with room-temperature ferromagnetic/ferrimagnetic ordering in addition to a high CO temperature.

**Acknowledgments**

This work was supported by the National Natural Science Foundation of China (50721001, 50831006) and the National Key Basic Research Program of China (2007CB925003).

Figure captions

Fig. 1 (a) Scheme of the current and magnetization measurements. (b) I-V curves in ab plane. The inset shows the measurement configuration. (c) The c-axis magnetization as a function of applied voltage pulses.

Fig. 2 Electrical switch of magnetization in $LuFe_2O_4$ at (a) 200 K and (b) 100 K. The curves marked with 0 T were measured as following：a 5 T magnetic field was applied along c axis to magnetize the sample, then the field is removed and the remanent magnetization was measured with time in zero magnetic field. The curves marked with 0.1 T and 5 T were measured in a constant 0.1 or 5 T field. The dotted lines refer to the positions where a 22 V (at 200 K) or 100 V (at 100 K) voltage pulse is applied.

Fig. 3 Temperature dependence of c-axis magnetization in 0.1 and 5 T with ZFC and FC processes. The green circles represent the initial and final magnetization states in 0.1 and 5 T shown in Fig. 2. The blue circle at 200 K marks the position where no switch-up of magnetization by voltage pulse is observed in a 5 T field. The cyan circle at 300 K marks the position where no magnetization change by applied voltage pulse is observed.

Fig. 4 Reversible switch of magnetic states at 200 K.

Fig. 5 Electrical manipulation of magnetization hysteresis loop at 100 K. The black square curve is the normal hysteresis loop measured after ZFC to 100 K. The red circle curve is the hysteresis loop measured with voltage pulses. The arrows mark the positions where a 100 V voltage pulse is applied.

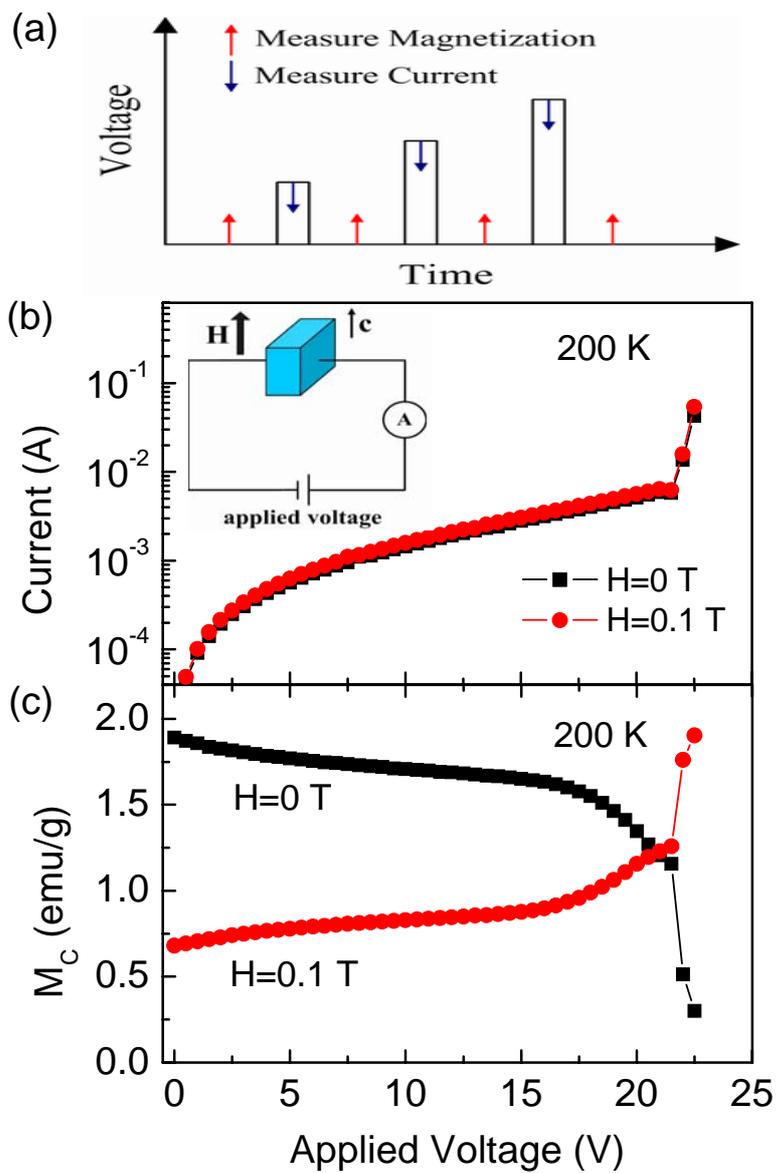

Fig. 2

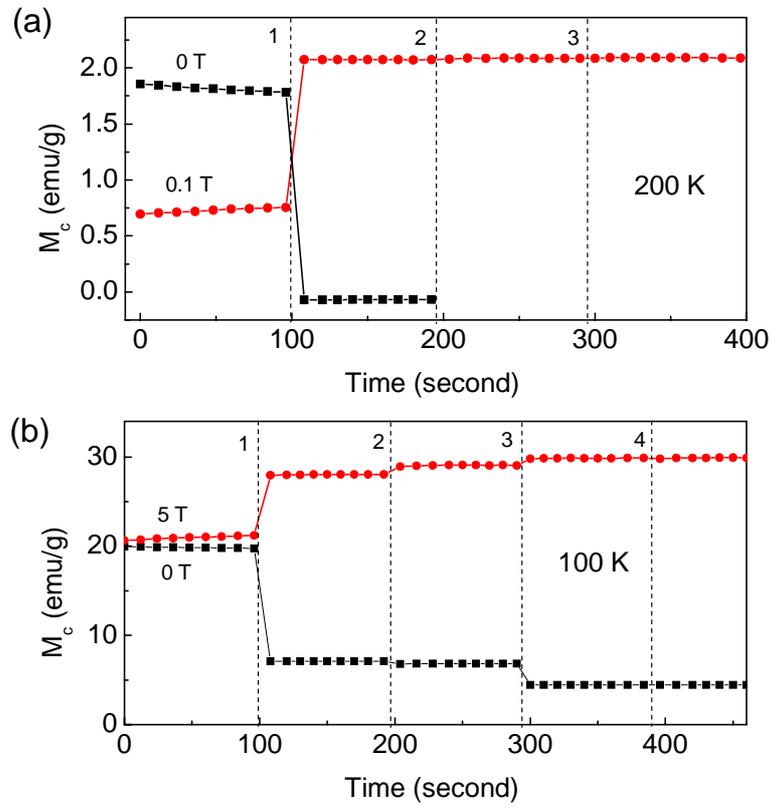

Fig. 3

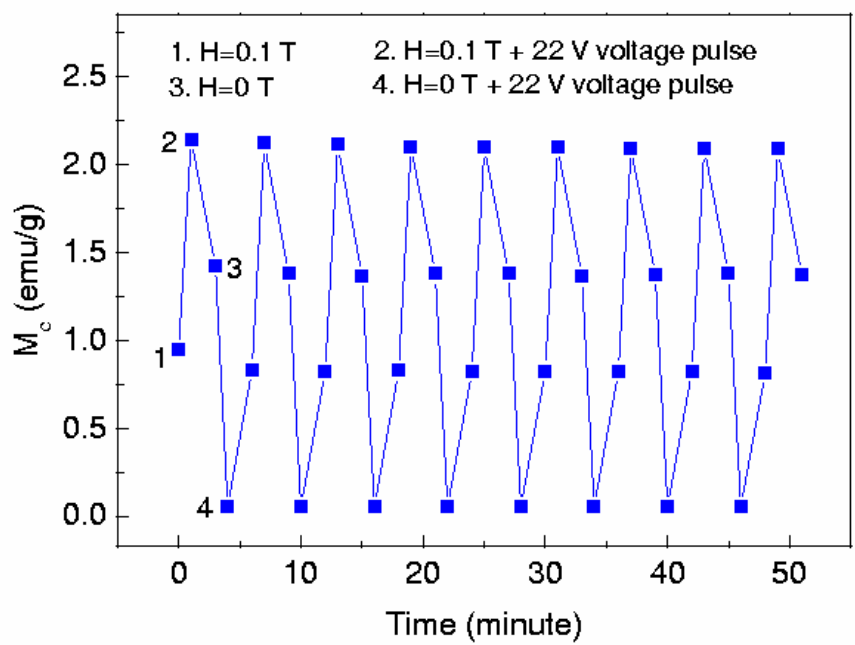

Fig. 5

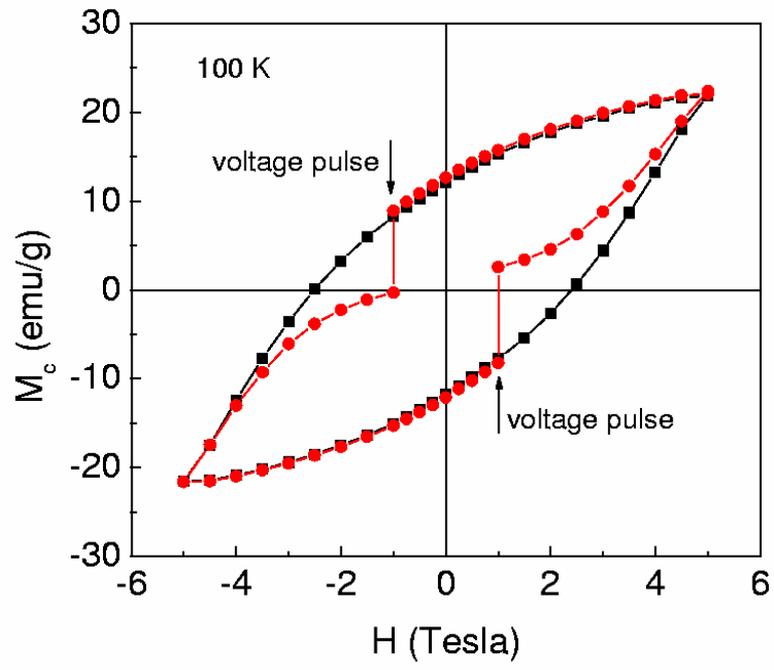